\documentclass[pdftex,twocolumn,epjc3]{svjour3} 

\RequirePackage[T1]{fontenc}

\smartqed  

\RequirePackage{graphicx}
\RequirePackage{mathptmx} 
\RequirePackage{flushend}
\RequirePackage[numbers,sort&compress]{natbib}
\RequirePackage[colorlinks,citecolor=blue,urlcolor=blue,linkcolor=blue]{hyperref}

\journalname{Eur. Phys. J. C}

\usepackage{graphicx}
\usepackage{dcolumn}
\usepackage{bm}
\usepackage{braket}
\usepackage{amsmath}
\usepackage{amsfonts}
\usepackage{here}
\usepackage{color}

\usepackage[small]{caption}
\usepackage[subrefformat=parens]{subcaption}
\begin{document}

\title{$\mathbb{Z}_N$ structure of deconfinement vacuum in SU($N$) Yang-Mills theory: emergence of Nambu-Goldstone mode in large-$N$ limit
}

\author{Yuto Nakajima\thanksref{e1,addr1}
        \and
        Hideo Suganuma\thanksref{e2,addr2} 
}

\thankstext{e1}{e-mail: nakajima.yuto.62s@st.kyoto-u.ac.jp}
\thankstext{e2}{e-mail: suganuma@scphys.kyoto-u.ac.jp}

\institute{Faculty of Science, Kyoto University, Kitashirakawa-oiwake, Sakyo, Kyoto 606-8502, Japan\label{addr1}
          \and
Department of Physics, Graduate School of Science, Kyoto University, Kitashirakawa-oiwake, Sakyo, Kyoto 606-8502, Japan\label{addr2}
}


\maketitle

\begin{abstract}
Using the Polyakov-loop effective action, we investigate the structure of spontaneously broken $\mathbb{Z}_N$ symmetry in the deconfinement vacuum in the SU($N$) Yang-Mills theory with finite $N$.
First, we examine the Polyakov-loop fluctuation around a $\mathbb{Z}_N$-broken vacuum 
and calculate the spatial correlation of its phase variable. 
We show that the phase variable of the Polyakov loop becomes a Nambu-Goldstone mode in the large-$N$ limit.  
%
Second, we estimate the global vacuum-to-vacuum transition rate in a finite-volume domain of the quark-gluon plasma.
Based on our estimation, we state that some threshold volume exists, a domain larger than which is stable, and vice versa.
Identifying the threshold as the lower bound of a stable center domain volume, we find the typical volume scale of center domains.
\end{abstract}

\section{Introduction}

Nowadays, quantum chromodynamics (QCD) is considered the fundamental theory of strong interaction. Among various phenomena generated by QCD, the color confinement is a crucial and fascinating non-perturbative aspect that characterizes physics at low energies as well as spontaneous chiral symmetry breaking. 
However, understanding the confinement remains a major challenge in theoretical physics, because of its non-Abelian properties and breakdown of the perturbative treatment in this regime. To overcome these theoretical difficulties, the lattice gauge theory was proposed ~\cite{1974Wilson,Kogut:1974ag}, and this method and its Monte Carlo simulation~\cite{Creutz:1980zw} enable us to treat confinement and to achieve significant success in hadron physics to date\cite{Rothe:1992nt}.

Color confinement, one of the most peculiar features of QCD, is lost at high temperatures, and the deconfined system, known as the quark-gluon plasma (QGP), has been well studied with great interest by many physicists in both theoretical and experimental sides. 
In fact, the QGP is believed to have existed in the early universe and has been created through relativistic heavy-ion collisions at RHIC \cite{STAR:2005gfr} and LHC-Alice experiment \cite{ALICE:2010suc, ALICE:2013mez}. 
As for the historical progress of lattice QCD, Polyakov and Susskind demonstrated using the strong coupling expansion that a confined system got deconfined at some temperature~\cite{1978Polyakov, 1979Susskind}. Using lattice QCD Monte Carlo calculations, Yaffe and Svetitsky investigated the QCD phase transition~\cite{1982Yaffe}, and Ogilvie demonstrated that the deconfinement transition is in the first-order in the Yang-Mills theory with three or more colors ~\cite{1984Ogilvie}.
A realistic SU(3) lattice QCD result with a large volume was presented by Columbia group \cite{Brown:1990ev} including dynamical u, d, and s-quarks with various masses. 
According to them, 
the QCD phase transition is in the strong first-order in the three-flavor chiral limit, 
it is in the weak first-order in the pure Yang-Mills limit with no dynamical quarks, 
and the QCD transition is crossover (no phase transition) in our real world with the physical quark mass.
This tendency was also confirmed by analyzing an $O(4)$ model in the same universality class \cite{Rajagopal:1992qz}.
Later, many lattice physicists have been quantitatively investigated the QCD transition temperatures in terms of chiral and deconfinement properties with large and fine lattices including small mass dynamical quarks \cite{Karsch:2001cy,Bazavov:2011nk}.
Recently, QCD phase transition at finite density has been also investigate in lattice QCD, with fighting at the notorious sign problem \cite{Fodor:2001au, Aarts:2009uq},
or using the strong-coupling expansion \cite{Fromm:2011qi}.

Recently, strong-coupling nature of QGP has been pointed out near the transition temperature $T_c$ from fluid-dynamical analyses for the QGP experimental data and lattice QCD studies. In fact, from the fluid-dynamical analysis of QGP, 
the shear viscosity $\eta$ per entropy density $s$ is found to be quite small as $\eta/s \simeq 0.1$, which is physically consistent with near perfect liquid \cite{Song:2010mg}, 
that is, a strong-coupling system, far from a quasi-free system. Some lattice QCD studies have also revealed the existence of $c$-$\bar c$ bound states even above $T_c$ 
below about $2T_c$ \cite{Asakawa:2003re,Iida:2006mv}
This strong-coupling system is called strongly-coupled QGP \cite{Shuryak:2004cy}, 
and some nonperturbative aspects are considered to remain even 
in the relatively low temperature QGP. 
This strong-coupling nature has been also supported from theoretical analyses with AdS/CFT correspondence for a supersymmetric version of the SU($N$) Yang-Mills theory \cite{Policastro:2001yc}: $\eta/s \simeq 1/(4\pi)$. 
In this way, even above the transition temperature $T_c$, the QCD system is expected to have strong-coupling properties near $T_c$, which needs some nonperturbative analysis.

The quark confinement in the Yang-Mills theory can be theoretically described using the Polyakov loop $L$, which was first introduced by Polyakov~\cite{1975Polyakov}. 
The thermal average of the traced Polyakov loop $\braket{\phi} = N^{-1}\braket{\mathrm{Tr} L}$ relates to 
the free energy of a static single quark $F_q$ as $\langle \phi \rangle \propto e^{-F_q/T}$
at the temperature $T$. 
If the thermal average is zero $\braket{\phi}=0$, the quark confinement occurs.
If it is non-zero $\braket{\phi}\neq 0$, the deconfinement is signified. 
Therefore, $\braket{\phi}$ serves as an order parameter for the quark confinement.

From a symmetry perspective, the deconfinement is interpreted as spontaneous symmetry breaking (SSB) of the global $\mathbb{Z}_N$ symmetry, which is the center of the gauge group SU($N$) \cite{1975Polyakov, 1978Polyakov}. 
In the lattice formalism, the Yang-Mills action, e.g. the plaquette action, is invariant under the global $\mathbb{Z}_N$ transformation, under which $\phi$ undergoes a transformation $\phi \mapsto z \phi \ (z \in \mathbb{Z}_N)$. 
For the lattice Yang-Mills theory, in the confinement phase, the $\mathbb{Z}_N$ invariance of $\braket{\phi}$ is respected, while in the deconfinement phase, the action retains the $\mathbb{Z}_N$ symmetry but $\braket{\phi}$ loses it at the state level. 
Thus, the deconfinement Yang-Mills vacuum has a nontrivial $\mathbb{Z}_N$ structure that spontaneously breaks. 

In this paper, we aim to analyze the structure of the deconfinement vacuum using the Polyakov loop effective model in the SU($N$) Yang-Mills theory. 
Svetitsky and Yaffe have derived a $d$-dimensional spin model that effectively describes the confinement from the $(d+1)$-dimensional Yang-Mills theory~\cite{1982Svetitsky}. 
This spin model has the $\mathbb{Z}_N$-symmetry and belongs to the same universality class as the $N$-state Potts model \cite{1952Potts}. 
The effective model is derived directly from the action of the Yang-Mills theory using the Migdal-Kadanoff renormalization group~\cite{1984Ogilvie, 1984Drouffe} or the strong coupling expansion~\cite{1982Polonyi, 1984Green}. 
Due to the strong coupling of the interactions in the quark-gluon plasma near the phase transition point, it seems reasonable to use the strong coupling approximation in the analysis.

Analytical studies of the deconfinement vacua at finite $N$ using the effective models have been conducted at the mean-field level in the literature such as the works by Matsuoka, Drouffe, or Ogilvie~\cite{1984Matsuoka, 1984Drouffe, 1984Ogilvie}. 
In this paper, we formulate an effective model incorporating non-uniformity and spatial fluctuations beyond the spatially uniform configuration. 
Using the effective action of the Polyakov-loop field $\phi(x)$,  
we analyze the properties of the Polyakov-loop fluctuation 
mainly in terms of its phase variable $\theta(x)$, 
while its amplitude $|\phi(x)|$ plays an important role in the confinement-deconfinement phase transition. 

In this paper, we consider any color number $N$ in the SU($N$) Yang-Mills theory.
In particular, the large-$N$ limit is interesting. 
The large-$N$ limit not only provides deeper insights into QCD but also contains interesting academic aspects such as the AdS/CFT correspondence \cite{
Maldacena:1997re, Witten:1998zw}, the dominant contribution of planar diagrams~\cite{1974tHooft}, or the quarkyonic phase in high-density QCD~\cite{2007McLerran}. 
Weiss presented the effective potential of the Polyakov loop and 
argued the large-$N$ behavior in the one-loop approximation 
for both the continuum and the lattice gauge theories \cite{Weiss:1980rj, Weiss:1981ev}.
Polony and Green formulated the Polyakov-loop effective model 
in a nonperturbative manner using the strong-coupling expansion \cite{1982Polonyi, 1984Green}, and 
Damgaard and Patk\'{o}s have exactly solved the effective model 
in the large $N$ limit~\cite{1986Damgaard}. 
Higher-order calculations by the character expansion have also been consistently performed by Bill\'{o} et al.~\cite{1994Billo}. 
While some studies assume $N$ to be infinite from the stage of constructing the effective model, we take the large-$N$ limit of the results obtained from the finite-$N$ effective model.
In this paper, we adopt a nonperturbative framework based on the strong-coupling expansion and restate the previous results in a more explicit form as described later.

Our first aim in this paper is to quantitatively evaluate the correlation length of the Polyakov loop phase $\theta(x)$.
The theory has $N$ different degenerate vacua at high temperatures, reflecting the spontaneously broken $\mathbb{Z}_N$ symmetry. 
One of these vacua is randomly chosen, and quantum and thermal effects cause local fluctuations around the vacuum. 
However, in the large-$N$ limit, we conjecture that the degenerate vacua get connected, and the global $\mathbb{Z}_N$ symmetry becomes an approximately continuous U(1). 
Then, the massive mode might become massless, as an extension of the Nambu-Goldstone theorem \cite{Nambu:1961tp, Nambu:1961fr, 1961Goldstone}.
In this paper, 
we investigate the SU(N) Yang-Mills theory with finite $N$ 
and the symmetry metamorphosis from $Z_N$ into $U(1)$ in large $N$, which was pointed out in Ref.~\cite{1986Damgaard}, 
and clarify its physical consequence such as emergence of 
the Nambu-Goldstone mode in an explicit manner.
We show in Sec. III and IV that the mass of the fluctuation along some direction vanishes for the Potts model and the Yang-Mills theory, respectively. 

\begin{figure}[thbp]
\centering
\includegraphics[width=86mm]{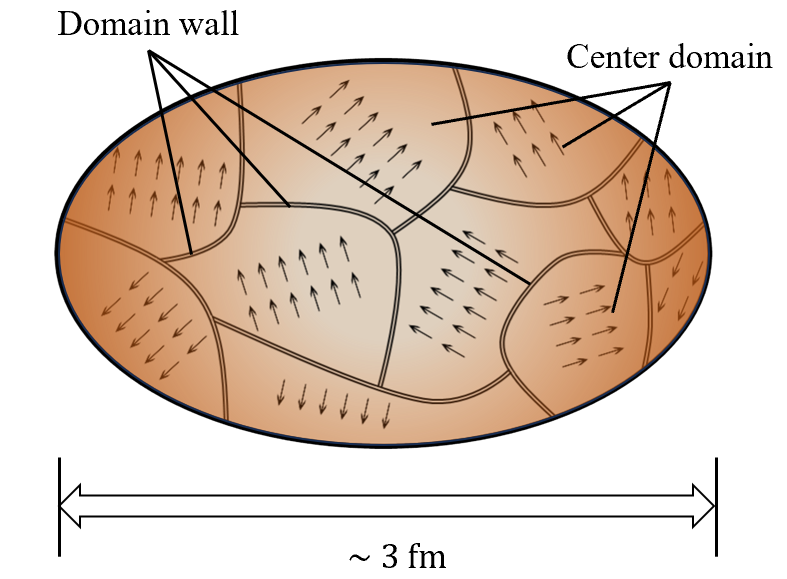}
\caption{The center domain of the quark-gluon plasma.
The arrows in the figure represent the phases of the vacuum expectation value of the Polyakov loop. 
Each domain, separated by the potential walls, is characterized by the vacuum configuration.}
\label{domainwall}
\end{figure}

Furthermore, another important aim of this paper is the effective description of stable \textit{center domains} in quark-gluon plasma.
The Polyakov loops prefer to stabilize in one of the potential minima and form local domains around one of the values $\phi = e^{i \frac{2\pi n}{N}}$ $(n = 0, 1, ..., N-1)$.
As a result, the plasma might have an inhomogeneous domain structure, where each domain is separated from the others by potential walls as shown in Fig.~\ref{domainwall}. 
The formation of these domains, called center domains, can be demonstrated by lattice simulation even with dynamical quarks~\cite{Borsnyi2010CoherentCD, STOKES2014341}.
Besides, some phenomenological arguments have been followed from the properties of the domains recently~\cite{PhysRevLett.110.202301}.
The center domains are expected to fluctuate due to quantum and thermal effects.
We estimate the timescale for the domains to remain in one potential minimum and demonstrate that the stability largely depends on their volumes: domains larger than some threshold are stable, and vice versa. 
Finally, we identify the volume threshold that determines its stability as the lower limit of the center domain volume.

The organization of this paper is as follows.
In Section \ref{sec2}, we summarize the general symmetry conversion from $\mathbb{Z}_N$ to U(1). 
In Sections \ref{sec3} and \ref{sec4}, we formulate the effective action and demonstrate that some fluctuations become massless in the large-$N$ limit in the Potts model and the SU($N$) Yang-Mills theory, respectively. 
In Section \ref{sec5}, we consider the global vacuum-to-vacuum transition in the center domain with a finite volume and evaluate its lifetime and stability as a function of the number of colors and the volume. 
Section \ref{sec6} is devoted to the conclusion.

\section{Symmetry conversion}
\label{sec2}

\begin{figure*}[htbp]
\centering
\includegraphics[width=129mm]{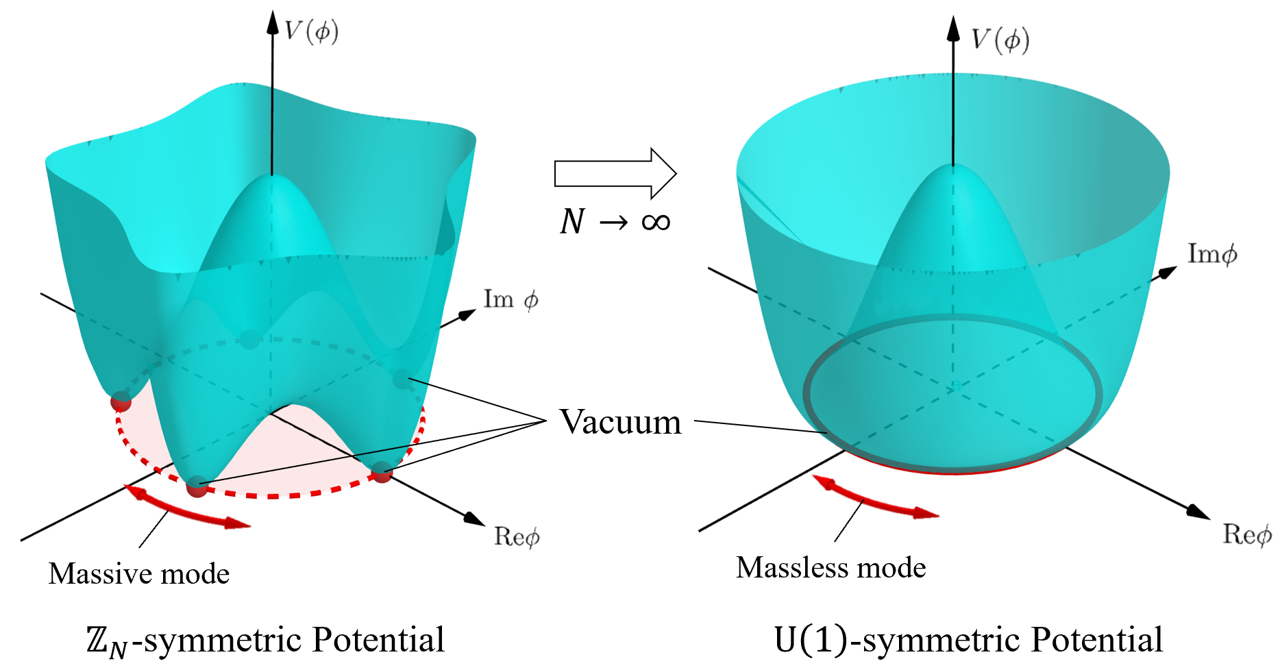}
\caption{In the large-$N$ limit, the $\mathbb{Z}_N$-symmetric potential is expected to become the U(1)-symmetric one, which results in the conversion from discrete symmetry breaking to continuous symmetry breaking. 
This generates a massless mode in the angular direction.}
\label{summary}
\end{figure*}

In this section, we summarize the basic properties of symmetry conversion from $\mathbb{Z}_N$ to U(1). 
We consider an arbitrary action $S[\varphi]$ involving a quantum field $\varphi$ with global $\mathbb{Z}_N$ symmetry: $S[\varphi] = S[z \varphi]$ $(z \in \mathbb{Z}_N)$. 
We suppose that the symmetry is spontaneously broken in terms of its vacuum expectation value $\langle \varphi \rangle \neq z \langle \varphi \rangle$. 
In other words, we consider a theory with $N$ different degenerate vacua. 
The problem is to determine whether the large-$N$ limit brings about qualitative changes in symmetry.

It is essential to bear in mind that $\mathbb{Z}_N$ is a discrete group. 
Nambu and Goldstone postulated that the spontaneous breaking of a global continuous symmetry results in the emergence of massless modes, the number of which corresponds to the number of broken degrees of freedom~\cite{Nambu:1961tp, Nambu:1961fr, 1961Goldstone}.
In this sense, the spontaneous breaking of $\mathbb{Z}_N$ symmetry does not necessarily result in the appearance of massless modes. 
However, massless modes can emerge in the large-$N$ limit because $\mathbb{Z}_N$ symmetry is expected to become approximately continuous U(1) symmetry, as depicted in Fig.~\ref{summary}.

Although this is \textit{a priori} speculation, we can demonstrate this type of conversion explicitly in some models. 
We achieve this by evaluating the correlation function of fluctuations in some direction and observing the behavior of the correlation length in the large-$N$ limit. 
The divergence of the correlation length (or the vanishing of the mode's mass) corresponds to the emergence of a Nambu-Goldstone mode. 
In the first part of this work, we demonstrate this conversion explicitly by using the $\mathbb{Z}_N$ Potts model and SU($N$) Yang-Mills theory.

\section{$\mathbb{Z}_N$ Potts model}
\label{sec3}

In this section, we describe the above transformation in statistical mechanics using the Potts model~\cite{1952Potts, 1982Wu}.
The model is defined by the Hamiltonian
\begin{align}
\beta H_{\mathrm{spin}} =
- \beta J \sum_{\braket{i, j}} (S_i^* S_j + \mathrm{c.c.}) 
\equiv - S^\dagger \hat J S,
\label{Hspin}
\end{align}
where $M$ spin variables $\{ S_i\} $ take the values $e^{i \frac{2 \pi n}{N}} \in \mathbb{Z}_N \ (n = 1, 2, ... , N) $ and the sum $\sum_{\braket{i, j}}$ is taken over all the nearest-neighbor spin variables.
$\beta$ is the inverse temperature of the system.
The Hermitian interaction coefficient matrix $\hat{J}$ is defined as follows:
\begin{align}
(\hat J)_{ij} = \left\{
\begin{array}{ll}
\beta J (>0) \quad &\text{($i$ and $j$ are the nearest neighbors)}\\
0  &\text{(otherwise)}
\end{array}
\right. 
\end{align}
It is obvious that (\ref{Hspin}) is invariant under the global $\mathbb{Z}_N$ transformation: $S_i \mapsto e^{i \frac{2\pi n}{N}}S_i$.

In the following, we show that the large-$N$ limit generates a novel massless mode in the $\mathbb{Z}_N$-broken low-temperature region.
The partition function for this system reads
\begin{align}
Z_{\mathrm{spin}}
= \mathrm{Tr} \ e^{S^{\dagger} \hat{J} S}.
\label{Zspin}
\end{align}
To derive the effective action involving a dynamical quantum field, we introduce an auxiliary complex scalar field $\phi = (\phi_1, ..., \phi_{M}) = (l_1 e^{i\theta_1}, l_2 e^{i\theta_2},... , l_{M} e^{i\theta_{M}})$ and convert (\ref{Zspin}) to the partition function with this dynamical field.
According to the method of Hubbard and Stratonovich, inserting the trivial path integral equation
\begin{align}
\int \mathcal{D}\phi \
e^{ 
-(\phi^{\dagger} - S^{\dagger} \hat{J} ) \hat{J}^{-1} (\phi - \hat{J} S)
}
=1
\end{align}
into (\ref{Zspin}), we obtain
\begin{align}
Z_{\mathrm{spin}}
&=
\int \mathcal{D}\phi \
e^{
- \phi^{\dagger} \hat{J}^{-1} \phi
}
\
\mathrm{Tr}\
e^ {
(
S^{\dagger} \phi + \phi^{\dagger} S
)
} 
\nonumber \\
&\equiv
\int \mathcal{D}\phi 
\exp \left[
- \phi ^{\dagger} \hat{J}^{-1} \phi
-
\sum_i V(\phi_i)
\right],
\label{Zspin2}
\end{align}
where
\begin{align}
V(\phi_i) \equiv
- \ln \left(
\sum_{n=1}^{N} 
\exp \left[2  l_i \cos \left(\theta_i - \frac{2 \pi n }{N}\right)
\right]
\right).
\label{V}
\end{align}
For simplicity, we omit the hat from $\hat{J}$ from now on.

 Taking the thermodynamical limit $M \rightarrow \infty$ and supposing the lattice spacing $a$ is small, we approximately reduce the sum over discrete indices $i$ to the integration over continuous variables as shown in \ref{analysisA}.
\begin{align}
\phi^{\dagger} J^{-1} \phi
\simeq
\frac{1}{36 J a^3}
\int \mathrm{d}^3 \boldsymbol{x}
\left(
|\nabla \phi (\boldsymbol{x})|^2 + \frac{6}{a^2} |\phi(\boldsymbol{x})|^2
\right)
\label{GradientExpansion}
\end{align}
We ignore the momentum higher-order terms to focus on the infrared region $|\boldsymbol{k}| \ll a^{-1}$ because now we are interested in the long-range correlations in the large-$N$ limit. 

Combining (\ref{Zspin2}), (\ref{V}), and (\ref{GradientExpansion}), we obtain an action for the three-dimensional complex scalar field $\phi$:
\begin{align}
&S_{\mathrm{spin}}[\phi(\boldsymbol{x})]
\nonumber \\
&=
\frac{1}{36Ja}
\int \mathrm{d}^3 \boldsymbol{x}
\Bigg(
|\nabla \phi(\boldsymbol{x})|^2+
\frac{6}{a^2}
|\phi(\boldsymbol{x})|^2
+
\frac{36J}{a^2}
V(\phi(\boldsymbol{x}))
\Bigg).
\label{Sspin}
\end{align}
The partition function reads $Z_{\mathrm{spin}}=
\int \mathcal{D}\phi
\exp [ - S_{\mathrm{spin}}[\phi(\boldsymbol{x})]]$.
In the low-temperature region,  the $\mathbb{Z}_N$ symmetry is spontaneously broken, i.e. the vacuum expectation value is nonzero: $\braket{\phi} \neq 0$.

Now we consider the fluctuations around a vacuum.
To focus on the fluctuation along the phase direction, we freeze the amplitude $|\langle\phi\rangle|=l_0$ and substitute $\phi(\boldsymbol{x}) = l_0 e^{i \theta(\boldsymbol{x})/l_0}$ into (\ref{Sspin}) to get
\begin{align}
    S_{\mathrm{spin}}
    &\sim 
\frac{1}{18Ja}
    \int \mathrm{d}^3 \boldsymbol{x}
\Bigg(
\frac{1}{2}
(\nabla \theta)^2
+ \frac{9J}{a^2}V''(l_0)
\theta^2
\Bigg) \nonumber \\
&\equiv
\frac{1}{18Ja}
\int \mathrm{d}^3 \boldsymbol{x}
\Bigg(
\frac{1}{2}
(\nabla \theta)^2+
\frac{m_{\mathrm{spin}}^2}{2}
\theta^2
\Bigg),
\end{align}
extracting the trivial constant.
Here we expand (\ref{V}) up to the second-order: $V(l_0 e^{i \theta(\boldsymbol{x})/l_0}) = V(l_0) + \frac{1}{2}V''(l_0)\theta^2 + \mathcal{O}(\theta^4)$
(Note that $\bullet'$ denotes the differentiation with respect to $\theta$.).
The invariance under the transformation $\theta \mapsto - \theta$ prohibits the odd-order terms.

Thus, the correlation function of $\theta$ reads
\begin{align}
\braket{\theta(\boldsymbol{x}) \theta(0)} 
=
\frac{\int\mathcal{D}\theta\ \theta(\boldsymbol{x})\ \theta(0)\ e^{-S_{\mathrm{spin}}}}
{\int\mathcal{D}\theta\ e^{-S_{\mathrm{spin}}}}
&\propto
\frac{1}{|\boldsymbol{x}|} e^{- m_{\mathrm{spin}} |\boldsymbol{x}|}.
\end{align}
We can see that the more explicit formula
\begin{align}
&\lim_{N \rightarrow \infty} m_{\mathrm{spin}}^2 \nonumber \\
&\propto \lim_{N \rightarrow \infty} 
\frac{\partial^2}{\partial\theta^2}
\ln \left(
\sum_{n=1}^{N} 
\exp \left[2  l_0 \cos \left(\frac{\theta}{l_0} - \frac{2 \pi n }{N}\right)\right]
\right)
\nonumber \\
&= 
\frac{\partial^2}{\partial\theta^2}
\ln \left(
\frac{1}{2 \pi} \int_0^{2 \pi} \mathrm{d}\varphi 
\exp \left[2  l_0 \cos \left(\frac{\theta}{l_0} - \frac{2 \pi n }{N}\right)\right]
\right)
=0
\end{align}
leads $m_{\mathrm{spin}}$ to vanish in the large-$N$ limit.
This clearly indicates that in the limit $\braket{\theta(\boldsymbol{x}) \theta(0)}$ becomes Coulomb-like i.e. a long-range correlation, or a Nambu-Goldstone mode, emerges.

\section{SU($N$) Yang-Mills theory}
\label{sec4}

In this section, we move to the SU($N$) Yang-Mills theory at finite temperature.
First of all, we briefly review the gauge theory on the lattice and formulation of the effective theory in the problem.
Without any dynamical fermions, the partition function is given by
\begin{align}
Z_{\mathrm{YM}} = \int \mathcal{D} U e^{-S_{\mathrm{YM}}[U]},
\end{align}
where $S_{\mathrm{YM}}[U]$ is the Wilsonian action~\cite{1974Wilson}:
\begin{align}
S_{\mathrm{YM}}[U] 
= -\frac{1}{2g^2} \sum_{\square} \mathrm{Re} \ \mathrm{Tr}\square_{\mu \nu}(n).
\label{plaquette}
\end{align}
The variables $U_{\mu}(n)$ and $\square_{\mu \nu}(n)$ represent the link variable and plaquette, respectively
\footnote{
Concretely, $U_{\mu}(n) = \exp[-i a g A_{\mu}(n)] \in {\rm SU}(N)$ and $\square_{\mu \nu}(n) = U_{\mu}(n) U_{\nu}(n +\hat{\mu}) \
U^{\dagger}_{\mu}(n + \hat{\nu}) \ U^{\dagger}_{\nu}(n)$. 
}.
The Haar measure on SU($N$) is represented by $\mathcal{D}U$, which is the product of the measure for all link variables. 
The sum over all possible plaquettes is denoted by $\sum_{\square}$.
To account for the finite temperature effect, we enforce periodic boundary conditions along the imaginary time $\tau$ direction with the period of inverse temperature $\beta = a N_{\tau}$ ($a$ is a lattice spacing). 
This is realized by setting $U_{\mu}(1, i) = U_{\mu}(N_{\tau}+1, i)$.

\begin{figure}[htbp]
\includegraphics[width=86mm]{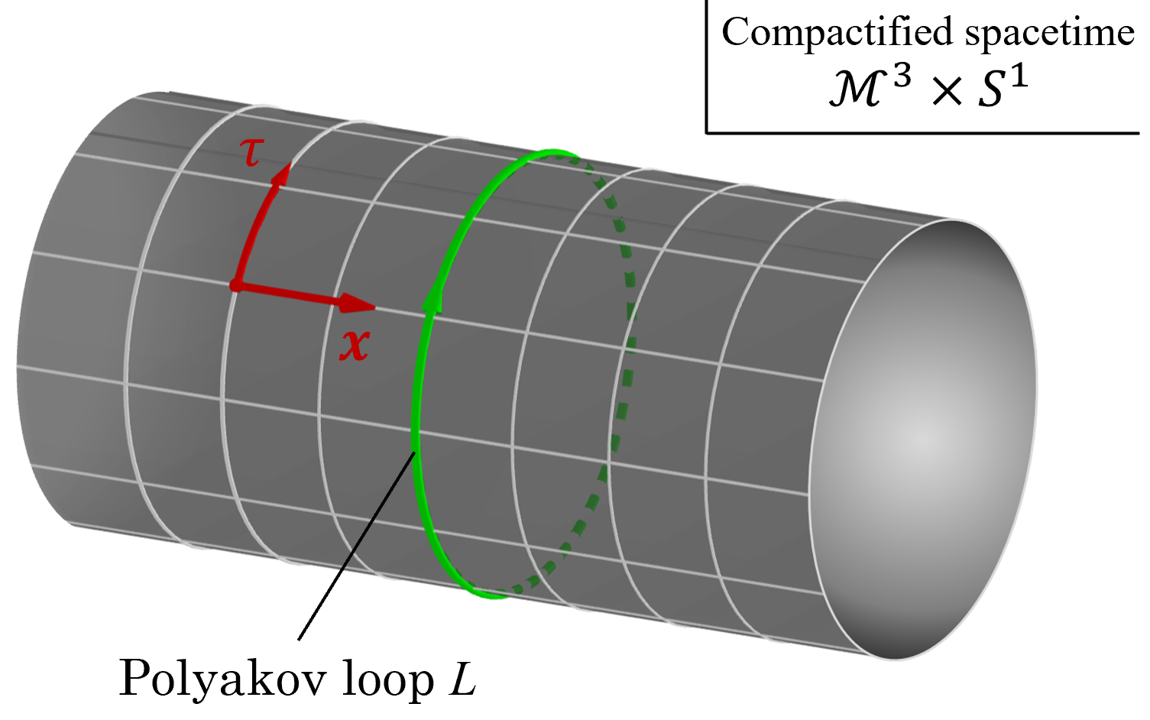}
\caption{The Polyakov loop on the lattice are shown in the figure.
At finite temperature, the Euclidean spacetime is periodic in the imaginary time $\tau$ direction.
After integrating out all the space-like link variables, the action depends only on the Polyakov loops under the strong coupling approximation. 
The thermal expectation value of the Polyakov loop serves as an order parameter for the deconfinement transition.}
\label{Spacetime}
\end{figure}

In the Yang-Mills theory at finite temperature, an order parameter of the confinement phase transition is the thermal average of the Polyakov loop 
\begin{align}
L_i = \prod_{i_{\tau}=1}^{N_{\tau}} U_{\tau} (i_{\tau}, i) \in {\rm SU}(N)
\end{align}
as shown in Fig. \ref{Spacetime}.
An effective action, whose dynamical variable is the Polyakov loop, should be invariant under the global $\mathbb{Z}_N$ transformation: $L_i \rightarrow z L_i \ (z \in \mathbb{Z}_N)$ and we expect the symmetry is spontaneously broken at high temperature.

The effective action can be formulated as 
\begin{align}
Z_{\mathrm{YM}}
&= \int \mathcal{D}U e^{-S_{\mathrm{YM}}[U]} \nonumber \\
&= \int \mathcal{D}U e^{-S_{\mathrm{YM}}[U]} 
\int \mathcal{D}\phi_i \ \delta \left[ \phi_i
- \frac{1}{N}\mathrm{Tr}(L_i)\right]\nonumber \\
&\equiv \int \mathcal{D}\phi \ e^{-S_{\mathrm{eff}}[\phi]}.
\label{ZYMnaive}
\end{align}
Although it is easy to see the structure of the symmetry, it is not clear how to calculate path integrals with such constraint conditions in (\ref{ZYMnaive}).
Therefore, we need to find another method to obtain a concrete form of effective action.

\subsection{Formulation}

The Polyakov loop effective action is obtained from (\ref{plaquette}).
This derivation is based on the Migdal-Kadanoff renormalization group~\cite{1984Ogilvie, 1984Drouffe} or the strong coupling expansion~\cite{1982Polonyi, 1984Green, Fromm:2011qi}.
From the perspective of strong coupling approximation, we integrate out all the space-like link variables and find
\begin{align}
    Z_{\mathrm{YM}} 
  =
    \int \left( \prod_{i_{\tau}=1}^{N_{\tau}} \prod_{i}\mathrm{d} U_{\tau}(i_{\tau}, i) \right)  \exp \left[ \lambda_{\mathrm{YM}}^{-N_{\tau}}\sum_{\braket{i, j}} \mathrm{Tr}( L_i ^{\dagger}) \ \mathrm{Tr}( L_j )\right]
    \label{ZYM00}
    \end{align}
up to the leading order, including ‘t~Hooft coupling $\lambda_{\mathrm{YM}}=g^2N$.
The sum $\Sigma_{\braket{i,j}}$ is taken over all the nearest-neighbor variables.

Here we take \textit{Polyakov gauge}, where all the time-like link variables are set to identity matrices except $U_{\tau}(1, i)$:
\begin{align}
U_{\tau}(i_{\tau}, i) = \left\{
\begin{array}{ll}
U_{\tau}(i_{\tau}, i) \qquad& \text{($i_{\tau}$ =1)}\\
1 & \text{($i_{\tau} = 2, 3, ... , N_{\tau}$)}
\end{array}
\right.
\end{align}
Under this gauge the integration $\int \mathrm{d}U_{\tau}(2,i)...\mathrm{d}U_{\tau}(N_{\tau},i)$ is trivial, and thus
\begin{align}
\prod_{i_{\tau}=1}^{N_{\tau}} \prod_{i}\mathrm{d} U_{\tau}(i_{\tau}, i) 
= \prod_{i} \mathrm{d}U_{\tau}(1,i)
\equiv \mathcal{D}L.
\end{align}
Now we can simplify (\ref{ZYM00}) as follows:
 \begin{align}
    Z_{\mathrm{YM}} 
    =
    \int \mathcal{D}L \exp \left[ e^{- \beta \sigma a} \sum_{\braket{i, j}} \mathrm{Tr}( L_i ^{\dagger}) \ \mathrm{Tr}( L_j ) \right]
\end{align}
with the string tension at zero temperature $\sigma =a^{-2} \ln \lambda_{\mathrm{YM}}$.

After the integral over SU($N$) is converted to the one over traced Polyakov loops $\phi_i = N^{-1} \mathrm{Tr}(L_i)$ with a Jacobian $\mathcal{H}^{(N)}(\phi_i)$, we find
\begin{align}
    Z_{\mathrm{YM}}
&= \int \mathcal{D}\phi \exp 
\left[N^2e^{- \beta \sigma a}  \sum_{\braket{i,j}} \phi_i^* \phi_j 
+\sum_i \ln \mathcal{H}^{(N)}(\phi_i)\right] \nonumber \\
&\equiv
\int \mathcal{D}\phi \exp 
\left[ - \phi^{\dagger} \hat{\mathcal{J}}^{-1} \phi
 + \sum_i \ln \mathcal{H}^{(N)}(\phi_i)\right],
\label{ZYM}
\end{align}
with $\phi = (\phi_1, \phi_2, ...)$ and a coefficient matrix $\hat{\mathcal{J}}^{-1}$ (we omit the hat for simplicity from now on).
(\ref{ZYM}) takes exactly the same form as (\ref{Zspin2}).
Through the same analysis as (\ref{GradientExpansion}), we obtain
\begin{align}
\phi^{\dagger} \hat{\mathcal{J}}^{-1} \phi
&\simeq
\frac{N^2 e^{-\beta \sigma a}}{a}
\int \mathrm{d}^3 \boldsymbol{x}
\left(
|\nabla \phi(\boldsymbol{x})|^2
- \frac{6}{a^2}|\phi(\boldsymbol{x})|^2
\right).
\label{Gradientexpansion2}
\end{align}
Such coarse graining of the short-wavelength modes, shown in \ref{analysisA}, leads to the effective action:
\begin{align}
&S_{\mathrm{YM}}[\phi(\boldsymbol{x})]
\nonumber \\
&=
C
\int \mathrm{d}^3 \boldsymbol{x}
\Bigg(
|\nabla \phi(\boldsymbol{x})|^2
-
\frac{6}{a^2}
|\phi(\boldsymbol{x})|^2
-
\frac{e^{\beta \sigma a}}{N^2 a^2}
\ln \mathcal{H}^{(N)} (\phi(\boldsymbol{x}))
\Bigg).
\label{SYM2}
\end{align}
The partition function reads $Z_{\mathrm{YM}} = \int \mathcal{D} \phi \ e^{-S_{\mathrm{YM}}[\phi(\boldsymbol{x})]}$ and $C \equiv a^{-1} N^2 e^{-\beta \sigma a}$.

Although we construct this model on a lattice, we can treat it as a continuous field model because we focus on the long-range correlations, specifically in the infrared region where $|\boldsymbol{k}| \ll a^{-1}$.
In other words, we assume that the lattice spacing is small enough compared to the typical wavelength of such correlations.

Next, we construct one of the important ingredients in (\ref{SYM2}), the Jacobian $\mathcal{H}^{(N)}(\phi)$.
When $N=2$ and $N=3$, it is known that the SU(2) Jacobian $\mathcal{H}^{(2)}$ and SU(3) Jacobian $\mathcal{H}^{(3)}$ reads~\cite{2005conrey}
\begin{align}
\mathcal{H}^{(2)}(\phi) &= 1 - \phi^2 \label{N=2}\\
\mathcal{H}^{(3)}(\phi) &= 1 - 6|\phi|^2 -3|\phi|^4 + 8 \mathrm{Re}\ \phi^3.\label{N=3}
\end{align}
They are expectedly invariant under the global $\mathbb{Z}_N$ transformation: $\phi \mapsto e^{i \frac{2 \pi n}{N}} \phi$.
Incidentally, the Jacobians (\ref{N=2}) and (\ref{N=3}) yield the second-order and the first-order deconfinement phase transition, respectively.
However, we now focus on describing fluctuations of the Polyakov loop \textit{phase}.
Although the Polyakov loop \textit{amplitude} is relevant for studying the deconfinement phase transition, it is not our present interest.

Finally, we examine the scenario when $N \geq 4$.
Since we have no exact form in these cases, we take the simplest form that preserves $\mathbb{Z}_N$ symmetry
\begin{align}
\mathcal{H}^{(N)}(\phi) = 1- b_2 |\phi|^2 - b_4 |\phi|^4 + b_N \mathrm{Re}\ \phi^N
\label{HN}
\end{align}
for our model according to Sannino's proposal~\cite{2005Sannino}
\footnote{
Although the $\mathbb{Z}_N$ invariance does not prohibit such higher-order terms as $|\phi|^6$, $|\phi|^8$, ..., or $\mathrm{Re}\phi^{2N}$, $\mathrm{Re}\phi^{3N}$, ..., we disregard these terms since their contributions are negligible (Note that $|\phi|<1$).
Besides, the term $\phi^N - \phi^{*N}$ is forbidden because it violates charge conjugation symmetry as pointed out in \cite{2005Sannino}.
}.
Certainly, this form might be unworkable to investigate the deconfinement phase transition because (\ref{HN}) yields the second-order transitions for even $N$'s, while they should be the first order.
However, it is not problematic as long as the discussion is limited to the deconfinement phase, where exclusively the Polyakov loop \textit{phase} is important.
Under such limitation of the effective model application, (\ref{HN}) is useful and workable.
For instance, this form is applied to the SU($N$) PNJL model in recent work~\cite{2012Buisseret} to investigate the chiral and confinement phase structure of SU($N$) Yang-Mills theory.
We also take (\ref{HN}) for our model from now on.

\subsection{Large-$N$ limit}

In this subsection, we confirm that (\ref{SYM2}) yields a massless mode in the large-$N$ limit explicitly.
We consider the fluctuations around a vacuum and investigate the phase $\theta(\boldsymbol{x})$ correlation function.
Freezing the amplitude $|\langle\phi(\boldsymbol{x})\rangle| = l_0$, we expand (\ref{HN}) in terms of $l_0^N(\ll1)$:
\begin{align}
\ln \mathcal{H}^{(N)}(l_0 e^{i \theta(\boldsymbol{x})/l_0})
\simeq
\ln{b}
- \frac{b_N l_0^N}{b} \left( 1- \cos \left( \frac{N}{l_0} \theta \right) \right),
\label{cos}
\end{align}
with $\phi(\boldsymbol{x}) = l_0 e^{i \theta(\boldsymbol{x})/l_0}$
(Note $b \equiv 1-b_2 l_0^2 - b_4 l_0^4 + b_N l_0^N$ and the upper bound condition $l_0 < 1$.).
Combining (\ref{SYM2}) and (\ref{cos}) and extracting the trivial constant, we obtain
\begin{align}
    S_{\mathrm{YM}}
    &\sim
    2C\int \mathrm{d}^3 \boldsymbol{x}
\left[
\frac{1}{2}
(\nabla \theta)^2 + V_{\mathrm{YM}}(\theta) 
\right],
\label{SYM3}
\end{align}
 where
 \begin{align}
&V_{\mathrm{YM}}(\theta) \equiv
\frac{m^2_{\mathrm{YM}}l_0^2}{N^2}
\left( 1 - \cos \left( \frac{N}{l_0}\theta \right)\right), \\
&m_{\mathrm{YM}} \equiv \sqrt{\frac{e^{\beta \sigma a}}{2 a^2}
\frac{b_N l_0^{N-2}}{ b}}.
\label{mYM}
\end{align}
 For a small $\theta$ fluctuation, we can focus on the vicinity of $\theta = 0$ and generate a mass term
\begin{align}
    S_{\mathrm{YM}}
    \simeq
2C\int \mathrm{d}^3 \boldsymbol{x}
\left[
\frac{1}{2}
(\nabla \theta)^2+
\frac{m_{\mathrm{YM}}^2}{2}
\theta^2
\right].
\label{SYM4}
\end{align}

Based on (\ref{SYM4}), the correlation function of $\theta(\boldsymbol{x})$ reads
\begin{align}
\braket{\theta(\boldsymbol{x}) \theta(0)}
=
\frac{\int \mathcal{D}\theta \ \theta(\boldsymbol{x}) \ \theta(0) \ e^{-S_{\mathrm{YM}}}}
{\int \mathcal{D}\theta e^{-S_{\mathrm{YM}}}} \propto
\frac{1}{|\boldsymbol{x}|} e^{-m_{\mathrm{YM}} |\boldsymbol{x}|}.
\end{align}
This mode gives a Yukawa-type spatial correlation with a range of $m_{\mathrm{YM}}^{-1}$. 
Taking into account (\ref{mYM}) and the upper bound constraint $l_0<1$, we conclude that 
\begin{align}
    \lim_{N \rightarrow \infty} m_{\mathrm{YM}} = 0,
\end{align}
namely
\begin{align}
\braket{\theta(\boldsymbol{x}) \theta(0)} \propto \frac{1}{|\boldsymbol{x}|}
\end{align}
in the large-$N$ limit.
This explicitly suggests that in the limit a Coulomb-type spatial correlation with an infinite range affecting the entire system, that is, the Nambu-Goldstone mode emerges.
Importantly, the mode is only massless in the limit while it remains massive for finite $N$.

Note that the mode does not propagate in spacetime dynamically, but rather represents a static and spatial long-range correlation. 
This is because the imaginary-time dependence has already been integrated out and real-time evolution is not included in the model. 
Since the mode originated from the fluctuation along the $\theta$ direction, it corresponds to a Nambu-Goldstone mode in a U(1)-symmetric quantum field theory.

In QCD, with three colors, the mass of this mode is approximately $m_{\mathrm{YM}} \simeq 2.1 \ \mathrm{GeV}$. 
This mass (\ref{mYM}) is calculated using a set of parameters: $a=0.4 \ \mathrm{fm}$, $\beta^{-1}=400 \ \mathrm{MeV}$, $\sigma = 1.0 \ \mathrm{GeV/fm}$, and $l_0=0.5$, as well as $b_2 = 6$, $b_4 = 3$, and $b_N = 8$ in (\ref{N=3}). 
Even though the mass of this mode is large compared to the typical mass scale of QCD, it becomes massless in the ideal large-$N$ limit.

\subsection{Quantum mechanical description}

In this subsection, we present the one-dimensional quantum mechanical action that describes the transition between the potential minima.
To estimate the transition rate, we must consider all possible paths connecting two vacua in the $\phi$-plane.
However, we assume that the dominant path is along the circumference $|\phi|=l_0$, which reduces the problem to a one-dimensional system.

Before we discuss the properties of (\ref{SYM3}) in detail, we must ensure that quantum corrections do not violate essential symmetries.
SSB due to quantum effects is a common phenomenon, as seen in the electroweak phase transition caused by quantum corrections, as derived by Coleman and Weinberg~\cite{1973Coleman}.
However, in the current system under study, it is demonstrated that quantum corrections up to one loop do not affect domain stability, and classical action suffices for physical considerations.

To verify this, we add a source term to (\ref{SYM3}): $Z_{\mathrm{YM}}[J] = \int \mathcal{D}\theta e^{-S_{\mathrm{YM}}+\theta\cdot J}$
\footnote{
$\theta \cdot  J=\int_V \mathrm{d}^3 \boldsymbol{x} \theta(\boldsymbol{x}) \  J(\boldsymbol{x})$
}
to find effective potential.
The generating functional of connected Green's functions
\begin{align}
W_{\mathrm{YM}}[ J] 
&\equiv -\ln Z_{\mathrm{YM}} [ J] \nonumber \\
&= S_{\mathrm{YM}}[\theta_{ J}] - \theta_{ J} \cdot  J + \frac{1}{2} \ln \det  S^{(2)}_{\mathrm{YM}}[\theta_{ J}]
\end{align}
 is defined for the saddle point configuration $\theta_{ J}$ that satisfies  
\begin{align}
     \frac{\partial S_{\mathrm{YM}}}{\partial \theta} =  J. 
 \end{align}
 $S^{(2)}_{\mathrm{YM}}[\theta]$ represents the second functional derivative of $S_{\mathrm{YM}}$.
By the Legendre transformation, we obtain $\Gamma _{\mathrm{YM}} (\langle \theta \rangle) 
\equiv W_{\mathrm{YM}} [ J] +  \langle \theta \rangle \cdot  J$:
\begin{align}
\Gamma _{\mathrm{YM}} (\braket{\theta}) 
&\simeq S_{\mathrm{YM}}(\braket{\theta}) + \frac{1}{2} \ln \det  S^{(2)}_{\mathrm{YM}}(\braket{\theta}) \nonumber \\
&= S_{\mathrm{YM}}(\braket{\theta})  + \frac{1}{2} \mathrm{Tr} \ln \Bigg( -\nabla^2 
 +m_{\mathrm{YM}}^2 \cos \left(\frac{N}{l_0} \braket{\theta} \right) \Bigg) \nonumber \\
&\simeq S(\braket{\theta}) 
+ \frac{V}{2} \int \frac{\mathrm{d}^3 \boldsymbol{k}}{(2 \pi)^3} \ln(k^2 + M_{\mathrm{YM}}(\braket{\theta})^2 )
\end{align}
up to one loop. 
Here $M_{\mathrm{YM}}(\braket{\theta})  \equiv m_{\mathrm{YM}} \cos^{1/2} \left( \frac{N}{l_0} \braket{\theta} \right)$.
Performing the integration, we obtain the effective potential:
\begin{align}
V _{\mathrm{eff}} (\braket{\theta}) 
&\simeq 
V_{\mathrm{YM}}(\braket{\theta}) 
- \frac{1}{2} \ \frac{\Gamma(-3/2)}{(4 \pi)^{3/2}} (M_{\mathrm{YM}}(\braket{\theta})^2)^{3/2} \nonumber \\
&=  V_{\mathrm{YM}}(\braket{\theta}) 
- \frac{1}{12\pi} M_{\mathrm{YM}}(\braket{\theta})^3 .
\label{Veff}
\end{align}
\begin{figure}[tbp]
\includegraphics[width=86mm]{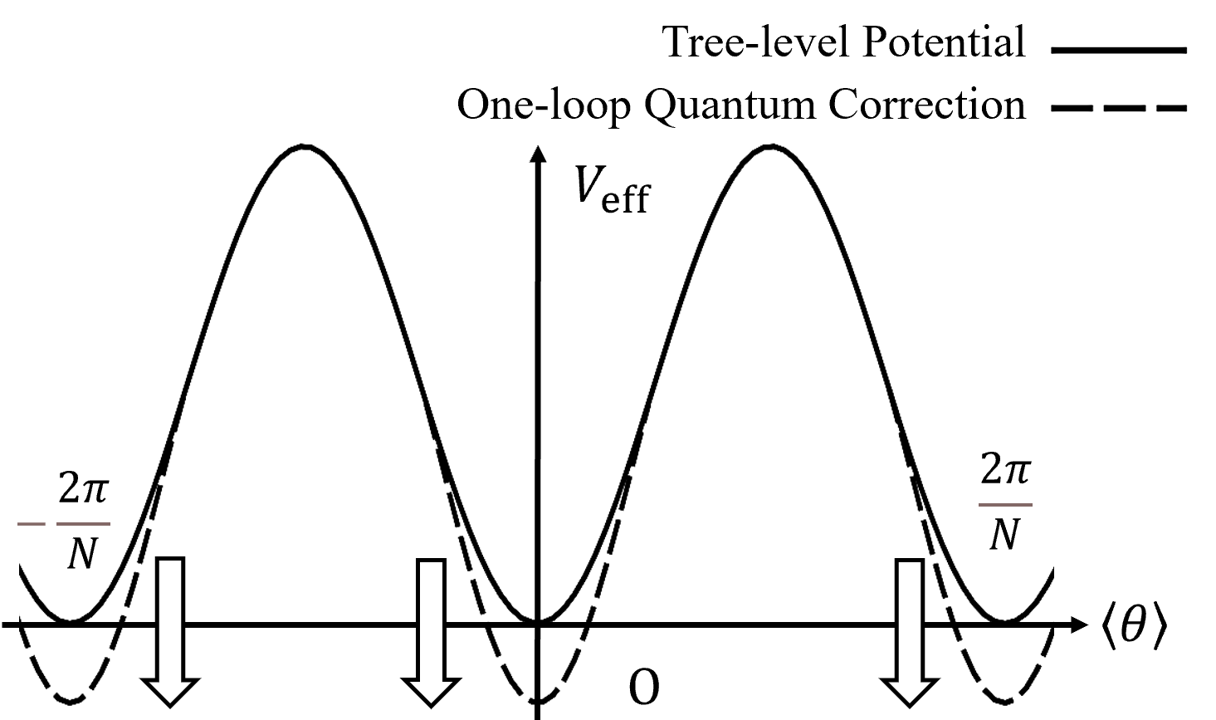}
\caption{The tree-level potential and the one-loop quantum correction are shown. 
Note that the locations of the minimum points are unchanged and the $\mathbb{Z}_N$ symmetry is preserved.}
\label{potential}
\end{figure}

The results shown in (\ref{Veff}) and Fig.~\ref{potential} clarify the two points: 
quantum corrections do not violate the $\mathbb{Z}_N$ symmetry of the action, and the contribution from quantum corrections vanishes faster than the tree-level term in the large-$N$ limit. 
As long as our discussion is limited to the systems with sufficiently large $N$, the quantum corrections do not play a significant role. 
Then, we safely ignore the quantum corrections and consider this problem at the tree level going forward.

In the following, we suppose that the field $\theta(\boldsymbol{x})$ has an imaginary-time dependence and treat $\theta$ as a real scalar field on a (1+3)-dimensional Euclidean spacetime.
We introduce the time derivative term with a parameter $\mathcal{Z}$ and define
\begin{align}
S_{\mathrm{QM}} 
=
\frac{2C}{a}\int \mathrm{d}\tau \int_V \mathrm{d}^3 \boldsymbol{x} 
\left[\frac{\mathcal{Z}}{2} \left(\frac{\partial \theta}{\partial \tau}  \right)^2 
+ \frac{1}{2} (\nabla \theta)^2 
+ V_{\mathrm{YM}}(\theta)\right]. \nonumber \\
\end{align}
The imaginary time formalism allows $\mathcal{Z}$ to be a non-trivial constant as space and time are no longer compatible variables in this framework. 
For the purpose of qualitative evaluation of the action, we put $\mathcal{Z}=1$ as an ansatz.

To analyze the homogeneous configuration in a finite domain, we suppose that $\theta(\tau, \boldsymbol{x})$ is homogeneous with respect to the spatial coordinate $\boldsymbol{x}$:
\begin{align}
S_{\mathrm{QM}} 
=
\frac{2CV}{a} \int \mathrm{d}\tau 
\left[\frac{1}{2} \left(\frac{\partial \theta}{\partial \tau}  \right)^2 
+ V_{\mathrm{YM}}(\theta)\right]. 
\label{SQM}
\end{align}
The action represents that of a quantum dynamical particle, and we interpret the vacuum-to-vacuum transition as the dynamics of some virtual particle following this action.

\section{Center Domain Volume} 
\label{sec5}

\begin{figure*}[htbp]
  \begin{minipage}[b]{0.48\linewidth}
      \centering
  \includegraphics[width=86mm]{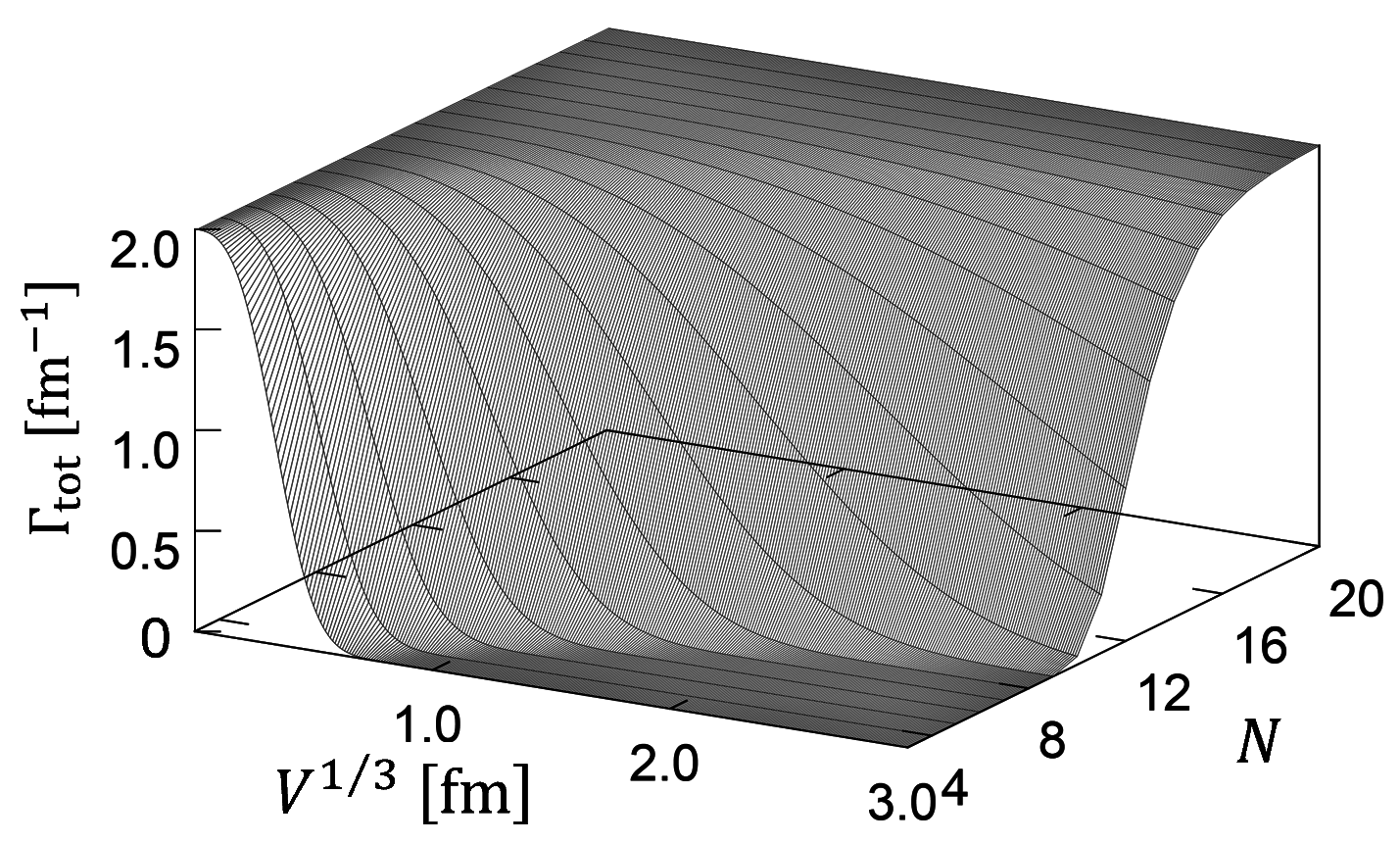}
  \end{minipage} %
  \begin{minipage}[b]{0.48\linewidth}
      \centering
  \includegraphics[width=86mm]{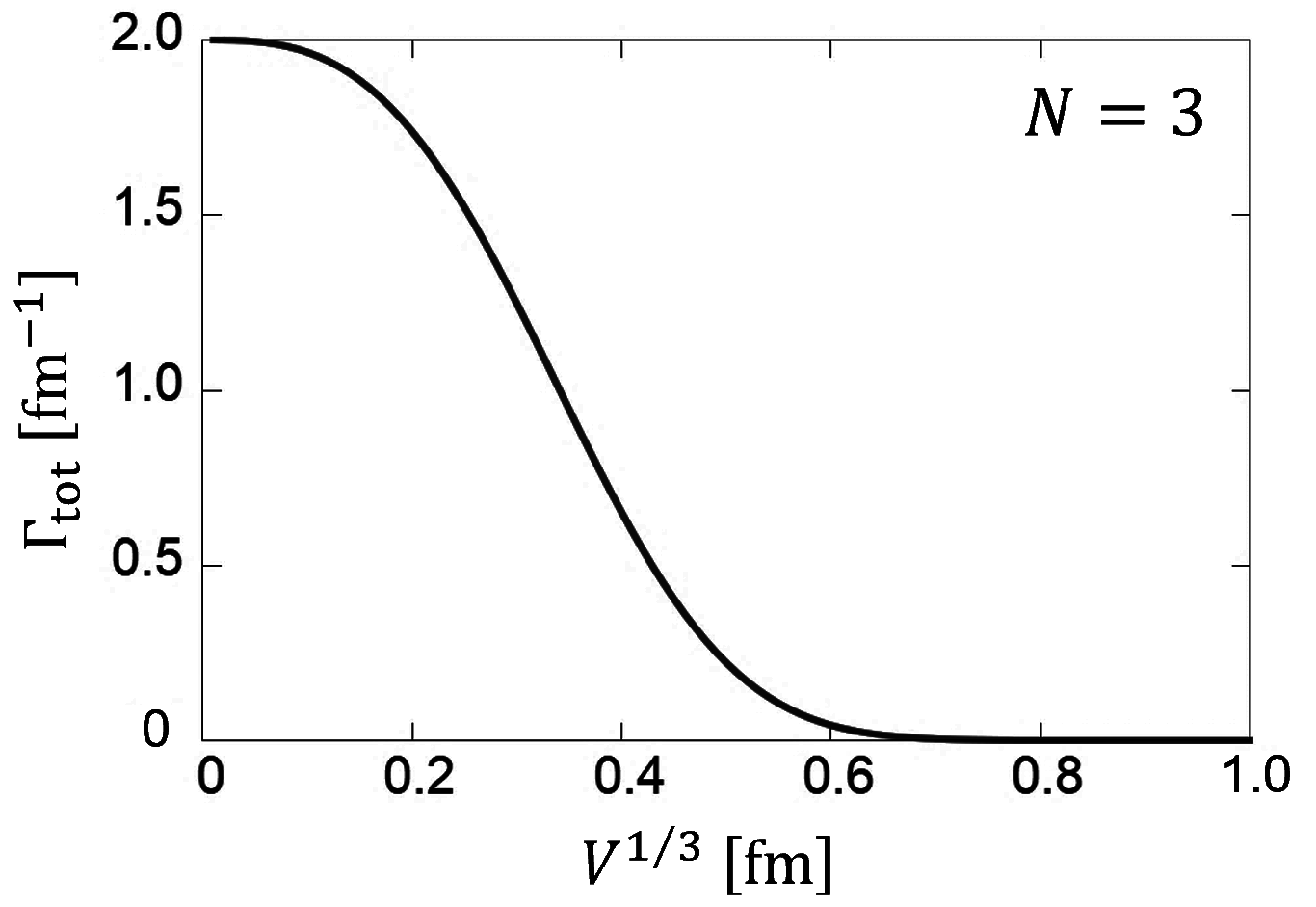}
  \end{minipage}
  \centering
  \caption{The figures show the total transition rate of the domain as a function of $V^{1/3}$ and $N$ (left panel) and for $N=3$ (right panel).
  These rates were calculated using the following parameter set: lattice spacing $a = 0.4 \ \mathrm{fm}$, temperature $T = 400 \ \mathrm{MeV}$, string tension at zero temperature $\sigma = 1.0 \ \mathrm{GeV/fm}$, $b_N/b = 7.52$ and the vacuum expectation value $l_0 = 0.5$.
Here we regard $b_N/b$ as the constant independent of $N$ and utilize $b_2 = 6$, $b_4 = 3$, and $b_N = 8$ in (\ref{N=3}) because the product of the parameter $b_N/b$ with $l_0^N$ always appears, and the $N$ dependence of the latter is expected to be more dominant in the large-$N$ limit. 
Therefore, it suffices to neglect the $N$ dependence of the former and focus on extracting the most significant $N$ dependence, $l_0^N$, to study the qualitative properties of the model.
The saturation of the rate in the large-$N$ region indicates that the transition is primarily driven by thermal processes, rather than quantum tunnelling.}
  \label{grapf_V,N}
\end{figure*}

 \begin{figure*}[htbp]
  \begin{minipage}[b]{0.48\linewidth}%
      \centering
  \includegraphics[width=86mm]{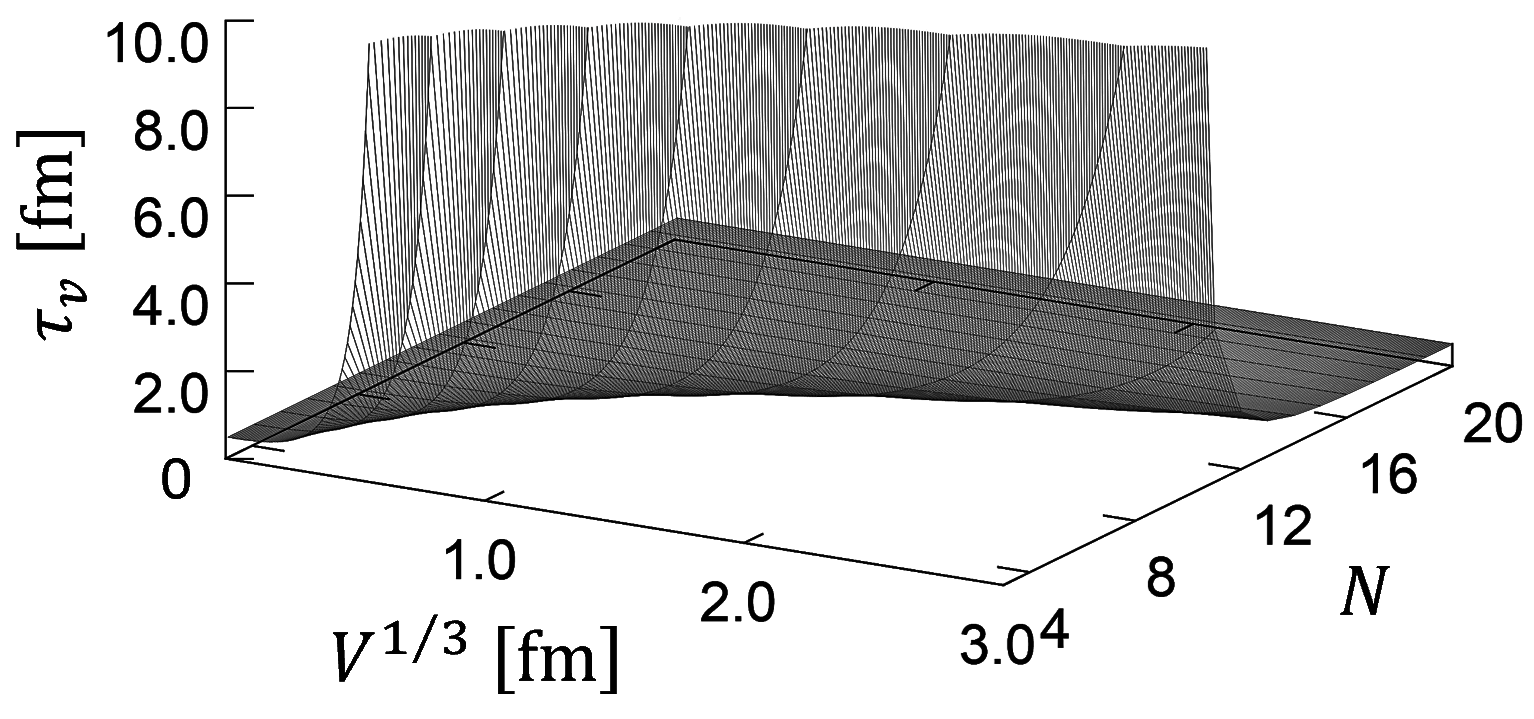}
  \label{lifetime}
  \end{minipage}%
  \begin{minipage}[b]{0.48\linewidth}%
      \centering
  \includegraphics[width=86mm]{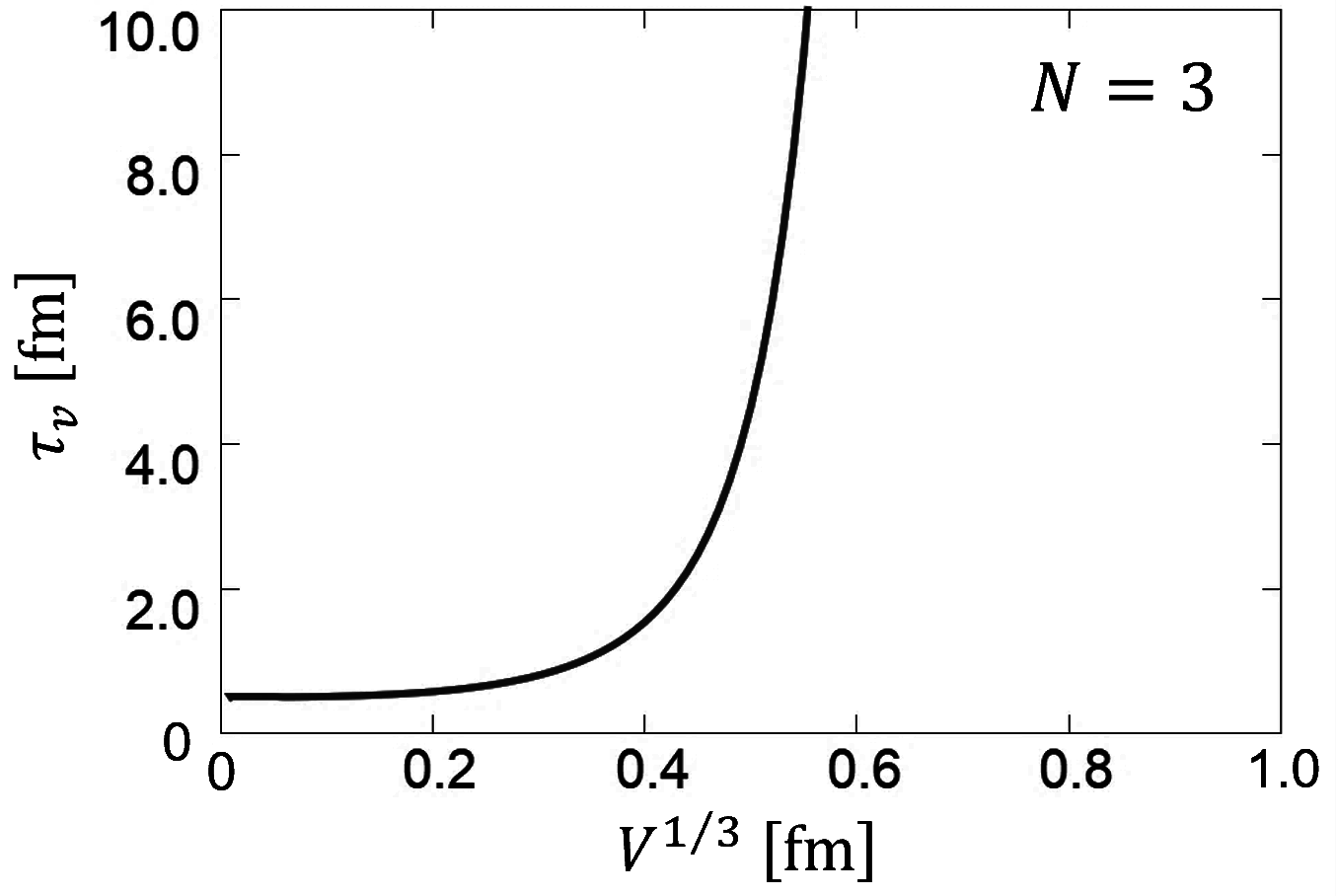}
  \end{minipage}
  \centering
  \caption{The figures show the lifetime of the domain as a function of $V^{1/3}$ and $N$ (left panel) and for $N=3$ (right panel). 
  These lifetimes were calculated using the same set of parameters as in Fig.~\ref{grapf_V,N}.
There is a specific line on which the lifetime sharply increases, dividing the $(V^{1/3}, N)$ plane into two distinct regions.
In the region of large $N$ and small $V$, the domain is unstable and easily transitions to another vacuum, while in the region of small $N$ and large $V$, the domain is stable.}
  \label{grapf2_V,N}
\end{figure*}

In this section, we estimate the lifetime of a domain and the center domain volume based on (\ref{SQM}).
For later convenience,  we define the coefficient in (\ref{SQM}) as follows:
\begin{align}
S_{\mathrm{QM}} 
\equiv
\int \mathrm{d}\tau 
\Bigg[\frac{M}{2} \left(\frac{\partial \theta}{\partial \tau}  \right)^2
+ \frac{V_0}{2}
\left( 1 - \cos \left( \frac{N}{l_0} \theta \right)\right)\Bigg],
\end{align}
where
\begin{align}
M(V,N) &\equiv \frac{2CV}{a} = \left( \frac{N}{a} \right)^2 \cdot 2V e^{-\beta \sigma a},\\
V_0(V,N) &\equiv \frac{2 l_0^2 m_{\mathrm{YM}}^2}{N^2a} \cdot 2CV
= \frac{2V}{a^4} \frac{b_N}{b} l_0^N
\label{V_0}
\end{align}
Note that in (\ref{V_0}) the $N$ dependence appears in $b_N/b$ and $l_0^N$, but the dominant contribution comes from the latter exponentiation.
Although $b_N/b$ is expected to vanish as $N \rightarrow \infty$, here we assume that the dependence is less dominant compared to $l_0^N$ for the later numerical calculation.

The transition between two adjacent wells can occur in two ways:
\begin{enumerate}
\item The thermal transition: a particle with energy $E \geq V_0(V, N)$ surmounts the potential barrier.
\item The quantum transition: a particle with energy $E < V_0(V, N)$ tunnels through the potential barrier.
\end{enumerate}
We define the transition rate from one well to the adjacent one per unit of imaginary time $\Gamma_{\mathrm{th}}(V, N)$ for case 1 and $\Gamma_{\mathrm{tun}} (V, N; E)$ for case 2.
The total transition rate per unit of imaginary time $\Gamma_{\mathrm{tot}}(V, N)$ is given by the sum of these two rates:
\begin{align}
\Gamma_{\mathrm{tot}}(V, N)  =\Gamma_{\mathrm{th}} (V, N) 
+\braket{\Gamma_{\mathrm{tun}}(V, N; E) },
\label{Gtot}
\end{align}
where $\braket{\bullet}$ denotes a thermal expectation value.

\subsection{On the ($V^{1/3}, N$)-plane}

In this subsection, we analytically calculate the transition rates $\Gamma_{\mathrm{tot}}(V,N)$ and estimate the lifetime of a domain $\tau_v(V,N) = 1/\Gamma_{\mathrm{tot}}(V,N)$.

$\Gamma_{\mathrm{th}} (V, N)$ can be evaluated easily:
\begin{align}
\Gamma_{\mathrm{th}}(V, N) 
= \frac{1}{\beta} 
\ \frac{\displaystyle \int_{V_0}^{\infty} d E e^{-\beta E}}
{\displaystyle \int_{0}^{\infty} d E e^{-\beta E}} 
= \frac{1}{\beta} \ e^{-\beta V_0(V, N)}.
\label{Gth}
\end{align}
Here $1/\beta$ is a typical frequency of thermal fluctuation called \textit{attempt frequency} in solid-state physics.

$\Gamma_{\mathrm{tun}}(V, N; E)$ can be evaluated using penetration rate per collision $P(V, N; E)$, or the \textit{Gamow factor}:
\begin{align}
\Gamma_{\mathrm{tun}}(V, N; E) = 2E \cdot P(V, N; E).
\end{align}
To obtain $P(V, N; E)$, a semi-classical estimate using the WKB approximation is sufficient:
\begin{align}
P(V, N; E) 
 = \exp \left[ -2 \int^{\frac{2\pi l_0}{N} - \theta_0}_{\theta_0} d\theta \sqrt{2 M (V(\theta) - E)} \right], 
 \label{WKB}
\end{align}
where $V(\theta_0) = E\ \left(0 \leq \theta_0 \leq \frac{l_0}{N} \pi \right)$. 
Using the expression, the thermal average of the transition rate can be calculated:
\begin{align}
\langle \Gamma_{\mathrm{tun}}(V, N; E) \rangle
 = \int _0^{V_0}  d E  \ 2E \cdot P(V, N; E)  e^{-\beta E}.
 \label{Gtun}
\end{align}

 Combining (\ref{Gth}) and (\ref{Gtun}), we evaluate (\ref{Gtot}) numerically using the parameter set as follows:  
lattice spacing $a = 0.4 \ \mathrm{fm}$, temperature $T = 400 \ \mathrm{MeV}$, string tension at zero temperature $\sigma = 1.0 \ \mathrm{GeV/fm}$, $b_N/b = 7.52$ and the vacuum expectation value $l_0 = 0.5$.
Note that here we regard $b_N/b$ as approximately the constant independent of $N$ and utilize $b_2 = 6$, $b_4 = 3$, and $b_N = 8$ in (\ref{N=3}).
This approximation is made because $b_N/b$ always appears as the product with $l_0^N$, and the $N$ dependence of the latter is expected to be more dominant in the large-$N$ limit. 
Therefore, it suffices to neglect the $N$ dependence of the former and to focus on the most significant $N$ dependence to demonstrate the qualitative properties of the model.

\begin{figure}[thbp]
\includegraphics[width=86mm]{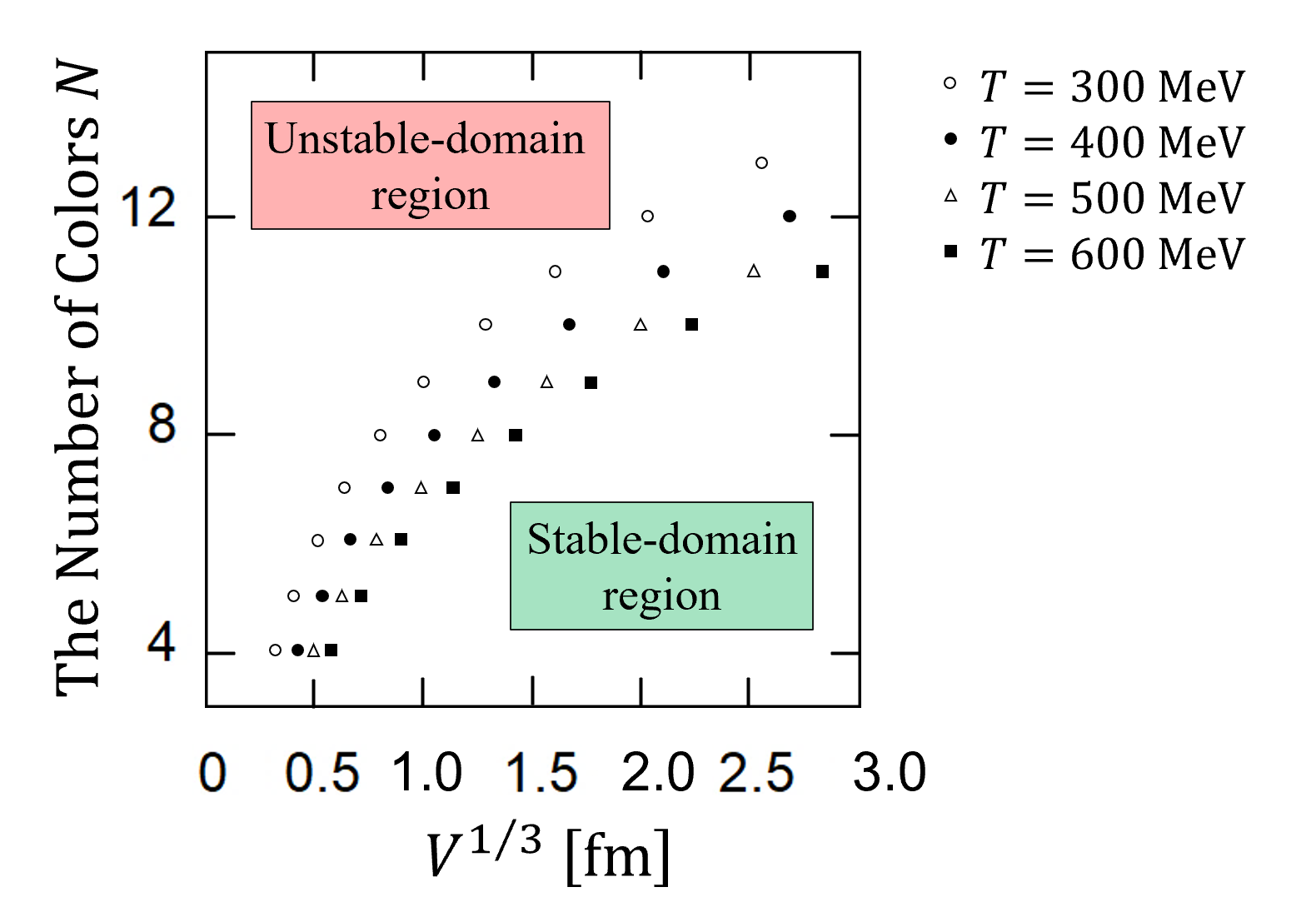}
\caption{The fuzzy boundaries between the unstable-domain and stable-domain regions are shown for different temperatures: $T$ = 300MeV, 400MeV, 500MeV, and 600MeV.
The domain's lifetime lasts for $1.0 \ \mathrm{fm}$ on the transition curves, which corresponds to the typical timescale of hadrons. 
Domains smaller than the transition thresholds, or unstable domains, would shrink and disappear, while those larger than the thresholds would be stabilized.}
\label{intercept}
\end{figure}

The figure in Fig.~\ref{grapf_V,N} displays the dependence of the total transition rate $\Gamma_{\mathrm{tot}}$ on $(V, N)$. 
On a specific curve, the transition rate sharply decreases. 
The lifetime $\tau_v$ is shown in Fig.~\ref{grapf2_V,N}. 
This figure exposes that the $(V^{1/3}, N)$-plane splits into two separate regions: the stable and unstable-domain regions. 
The boundary between these two phases is a fuzzy crossover, defined by the rapid rise in lifetime. 
The “critical curve," where the lifetime intersects 1.0 $\mathrm{fm}$, is illustrated in Fig.~\ref{intercept} for $T=$ 300MeV, 400MeV, 500MeV, and 600MeV.

It is important to note that the crossover curves are also dependent on the lattice spacing, which is one of the parameters of the model. 
Considering the strong coupling approximation and the asymptotic freedom, we set the lattice spacing $a$ to 0.4 $\mathrm{fm}$, which is larger than the typical length scale of hadrons but smaller than that of the modes studied.

Our findings provide insights into the structure of center domains and domain walls in the quark-gluon plasma. 
In high-energy heavy-ion collision experiments, the system is divided into thousands of small center domains with different vacuum configurations.
Considering that the configuration of the domain more minor than the “critical volume" is likely to decay, the typical lower bound of the center domain volumes corresponds to the “critical volume."

\subsection{Some extreme cases}

In addition to the above results for finite $N$ and $V$, we consider the domain stability in such extreme cases as $V \rightarrow \infty$ and the large-$N$ limit.

Both $V_0(V, N)$ and $M(V, N)$ diverge in the $V \rightarrow \infty$ limit, and thus $\Gamma_{\mathrm{th}}(V,N)$ vanishes.
In addition, using $\theta_0 = 0$ that follows from $V_0 \gg E$, (\ref{WKB}) becomes
\begin{align}
P(V, N; E) 
&\simeq \exp \left[ -2 \int^{\frac{2\pi l_0}{N}}_0 d\theta \sqrt{2 M V(\theta)} \right] \nonumber \\
&= \exp \left[ - \frac{16 \sqrt{M V_0}}{N}\right].
\end{align}
Then, the thermal transition rate also vanishes:
\begin{align}
\lim_{V\rightarrow \infty}\langle \Gamma_{\mathrm{tun}}(V, N; E) \rangle
&= \lim_{V\rightarrow \infty} 2\beta \ P(V, N; 0)\int ^{\infty}_0 d E \ E  e^{-\beta E} 
\nonumber \\ &= 0.
\end{align}
Therefore, we conclude that vacuum-to-vacuum transition is perfectly suppressed when $V \rightarrow \infty$:
\begin{align}
\lim_{V \rightarrow \infty}\Gamma_{\mathrm{tot}}(V,N) = 0.
\end{align}
In other words, the infinitely large domains are stable: 
\begin{align}
\lim_{V \rightarrow \infty}\tau_v(V,N)= \infty.
\end{align}
Note that we can also derive the irreducibility using the instanton method in \ref{instanton}.

In the large-$N$ limit, $V_0(V, N)$ vanishes while $M(V, N)$ diverges, which results in
\begin{align}
\lim_{N \rightarrow \infty}\Gamma_{\mathrm{th}}(V,N) = \frac{1}{\beta}
\end{align}
and considering $0 < E \leq V_0$,
\begin{align}
\lim_{N \rightarrow \infty} \braket{\Gamma_{\mathrm{tun}}(V, N; E)}
= \Gamma_{\mathrm{tun}} (E=0)
=0.
\end{align}
Therefore, the dominant contribution comes from the thermal effect rather than the tunneling.
Consequently, a particle is completely free from the potential, and the lifetime of the domain is equivalent to the thermal fluctuation:
\begin{align}
\lim_{N \rightarrow \infty}\tau_v(V,N) = \beta.
\end{align}

\section{Conclusion}
\label{sec6}

In this paper, we have investigated the non-trivial $\mathbb{Z}_N$ structure of the deconfinement vacuum in SU($N$) Yang-Mills theory. 
We expand the known effective action of quenched QCD to the SU($N$) Yang-Mills theory, considering spatial dependence.

In the first section, we have developed the Polyakov-loop effective action for the theory with a finite number of colors. 
We examine the correlation function of the Polyakov loop and derive the fluctuation mass as a function of the color number, based on the model. 
We have also analyzed the global symmetry structure metamorphosis in the large-$N$ limit, focusing on the fluctuation of the Polyakov loop phase, and have found that the mode becomes  massless in the large $N_c$ limit, which can be regarded as an extension of the Nambu-Goldstone theorem. 
This confirms that this limit changes the $\mathbb{Z}_N$-symmetric theory into a U(1)-symmetric theory.

In the second section, we have investigated the global $\mathbb{Z}_N$ structure in the finite-volume quark-gluon plasma. 
We model the global transition as a one-dimensional movement of a particle to compute the transition probability between different degenerate vacua. 
We consider both the thermal and the quantum effects and calculate the lifetime of a domain until it decays. Specifically, we show that the lifetime diverges in the infinite volume limit and vanishes in the large-$N$ limit.

In the last section, we have studied the typical volume scale of one of the center domains in the quark-gluon plasma as a function of the color number and volume. 
We discover that the $(V^{1/3}, N)$-plane can be divided into two regions: stable-domain and unstable-domain regions. 
This indicates that a center domain is stable if its volume exceeds a certain threshold, while a center domain with a volume below it is unstable. 
We identify the threshold as the lower bound of a stable center domain volume.

In conclusion, we have provided 
a useful description of the $\mathbb{Z}_N$ structure 
in the deconfinement vacuum of SU($N$) Yang-Mills theory  
based on the Polyakov-loop effective action.

In this paper, to get the theoretical outline of $\mathbb{Z}_N$ structure on the QCD vacuum, we have dealt with the Yang-Mills theory without dynamical quarks as an idealized limit of QCD. 
As an important next step, we aim to include dynamical quarks 
and investigate the quark effect of the above arguments.
While the $\mathbb{Z}_N$ symmetry is exact in the pure Yang-Mills theory, the quark action breaks it explicitly \cite{Rothe:1992nt}, and the vacuum with real $\braket{\phi}$ is energetically favored 
in quark-gluon plasma. 
This quark effect would bring some modification to the $\mathbb{Z}_N$ domain structure of quark-gluon-plasma, 
which is expected to depend significantly on the quark mass. 
It is also interesting 
to investigate the $\mathbb{Z}_N$ domain structure 
in the finite baryon-density quark-gluon plasma. 

As a technical improvement, 
beyond the strong-coupling expansion in this paper, 
it is also important and desired to perform 
lattice QCD Monte-Carlo simulations and 
to numerically verify our results.

\begin{acknowledgements}
H.S. is supported in part by the Grants-in-Aid for
Scientific Research [19K03869] from Japan Society for the Promotion of Science.
\end{acknowledgements}

\appendix

\section{Coarse graining of nearest-neighbor interactions}
\label{analysisA}

In this appendix, the coarse graining of the interaction on the lattice in (\ref{GradientExpansion}) and (\ref{Gradientexpansion2}) is reviewed.

\subsection{For $\mathbb{Z}_N$ Potts model}
\label{analysisA1}

First, in the momentum space, the left-hand side  in (\ref{GradientExpansion}) reads 
\begin{align}
\sum_{i, j} \phi_i^* J_{ij}^{-1} \phi_j  
=
M^{-1/2}
\sum_{\boldsymbol{k}}
\frac{1}{\tilde{J}_{\boldsymbol{k}}}
|\tilde{\phi}_{\boldsymbol{k}}|^2.
\end{align}
where $\phi_i  = M^{^-1/2}
 \sum_{\boldsymbol{k}}\tilde{\phi}_{\boldsymbol{k}} 
 e^{i \boldsymbol{k} \cdot \boldsymbol{x}_i}$ and $J_{ij} = M^{-1/2}
\sum_{\boldsymbol{k}} \tilde{J}_{\boldsymbol{k}}
e^{i \boldsymbol{k} \cdot \boldsymbol{r}}$ 
$(\boldsymbol{r} = \boldsymbol{x_i - x_j})$.  
Using the matrix product $\sum_j J_{ij} J^{-1}_{jk} = \delta_{ik}$, we find $(\tilde{J}^{-1})_{\boldsymbol{k}} 
= (M \tilde{J}_{\boldsymbol{k}})^{-1}$.

 Taking the thermodynamical limit $M \rightarrow \infty$ and supposing the lattice spacing $a$ is small compared to the typical correlation length, we approximately reduce the sum over discrete indices $i$ to the integration over continuous variables.
Namely, we define $\phi(\boldsymbol{x}) = \phi_i (J(\boldsymbol{r}) = J_{ij} )$ and $\tilde{\phi}(\boldsymbol{k}) = a^3 M^{1/2} \tilde{\phi}_{\boldsymbol{k}} (\tilde{J}(\boldsymbol{k}) = a^3 M^{1/2} \tilde{J}_{\boldsymbol{k}})$ to reproduce the Fourier transformation of the continuous variables $\phi(\boldsymbol{x}) 
= \int \frac{\mathrm{d^3} \boldsymbol{p} }{(2\pi)^3}
\tilde{\phi}(\boldsymbol{k}) 
e^{i \boldsymbol{k} \cdot \boldsymbol{x}}
$ and $\tilde{\phi}(\boldsymbol{k}) 
= \int \mathrm{d^3} \boldsymbol{x} 
\phi(\boldsymbol{x}) 
e^{-i \boldsymbol{k} \cdot \boldsymbol{x}}$. 

Since the nearest-neighbor interaction can be written in the form $J(\boldsymbol{r}) = J a^3 \sum_{i=1}^{6} \delta^3 (\boldsymbol{r} - \boldsymbol{a}_i)$, we obtain 
\begin{align}
\tilde{J}(\boldsymbol{k}) = Ja^3 \sum_{i=1}^6 e^{-i \boldsymbol{k}\cdot \boldsymbol{a}_i} = Ja^3 (6 - a^2 \boldsymbol{k}^2 ) + \mathcal{O}(\boldsymbol{k}^4).
\end{align}
$\{\boldsymbol{a_i}\}$ denotes a set of the nearest grid points from the origin.

Thus we find
\begin{align}
M^{-1/2}
\sum_{\boldsymbol{k}}
\frac{1}{\tilde{J}_{\boldsymbol{k}}}
|\tilde{\phi}_{\boldsymbol{k}}|^2
&\simeq 
\int \frac{\mathrm{d}^3 \boldsymbol{k}}{(2 \pi)^3}
\frac{1}{\tilde{J}(\boldsymbol{k}) }
|\tilde{\phi}(\boldsymbol{k})  |^2
\nonumber \\
&\simeq
\frac{1}{6 J a^3}
\int \frac{\mathrm{d}^3 \boldsymbol{k}}{(2 \pi)^3}
\left(
1 + 
\frac{a^2}{6}
\boldsymbol{k}^2
\right)
|\tilde{\phi}(\boldsymbol{k})  |^2. \nonumber \\
&=
\frac{1}{36 J a^3}
\int \mathrm{d}^3 \boldsymbol{x}
\left(
|\nabla \phi (\boldsymbol{x})|^2 + \frac{6}{a^2} |\phi(\boldsymbol{x})|^2
\right)
\end{align}
In the second line, we expand $\tilde{J}(\boldsymbol{k})$ in terms of $\boldsymbol{k}$. 

\subsection{For SU($N$) Yang-Mills theory}

In the same way as discussed above, we find that the left-hand side in (\ref{Gradientexpansion2}) becomes
\begin{align}
\sum_{i, j}\phi_i^* \hat{\mathcal{J}}_{ij}^{-1} \phi_j 
=
M^{-1/2}
\sum_{\boldsymbol{k}}
\frac{1}{\tilde{\mathcal{J}}_{\boldsymbol{k}}}
|\tilde{\phi}_{\boldsymbol{k}}|^2
\end{align}
In this case, the nearest-neighbor interaction function $\mathcal{J}^{-1}(\boldsymbol{r}) = -N^2 a^3 e^{-\beta \sigma a}\sum_{i=1}^{6} \delta^3 (\boldsymbol{r} - \boldsymbol{a}_i)$ can be written as follows:
\begin{align}
\tilde{\mathcal{J}}^{-1}(\boldsymbol{k}) 
= -N^2 a^3 e^{-\beta \sigma a} (6 - a^2 \boldsymbol{k}^2 ) + \mathcal{O}(\boldsymbol{k}^4).
\end{align} 

Therefore, we find the final form 
 \begin{align}
M^{-1/2}
\sum_{\boldsymbol{k}}
\frac{1}{\tilde{\mathcal{J}}_{\boldsymbol{k}}}
|\tilde{\phi}_{\boldsymbol{k}}|^2
&\simeq 
\int \frac{\mathrm{d}^3 \boldsymbol{k}}{(2 \pi)^3}
\frac{1}{\tilde{\mathcal{J}}(\boldsymbol{k}) }
|\tilde{\phi}(\boldsymbol{k})  |^2
\nonumber \\
&\simeq
- \frac{6 N^2 e^{-\beta \sigma a}}{a^3}
\int \frac{\mathrm{d}^3 \boldsymbol{k}}{(2\pi)^3}
\left(1 - 
\frac{a^2}{6} \boldsymbol{k}^2 
\right) |\phi(\boldsymbol{k})|^2\nonumber \\
&\simeq
\frac{N^2 e^{-\beta \sigma a}}{a}
\int \mathrm{d}^3 \boldsymbol{x}
\left(
|\nabla \phi(\boldsymbol{x})|^2
- \frac{6}{a^2}|\phi(\boldsymbol{x})|^2
\right).
\end{align}

\section{Tunnelling estimation with instantons}
\label{instanton}

In this appendix, we see that different vacua become irreducible as $N \rightarrow \infty$ using the\textit{ instanton }method.
For later convenience, we transform $\theta \mapsto \theta - (l_0/N)\pi$ and rewrite (\ref{SQM}):
\begin{align}
S_{\mathrm{QM}} 
&=
\int \mathrm{d} \eta  \left[ \frac{1}{2} \left( \frac{\partial \Theta }{\partial \eta } \right)^2 + \frac{1}{G^2} (1 + \cos(G \Theta )) \right] \nonumber\\
&\equiv
\int \mathrm{d} \eta  \ \mathcal{L}_{\mathrm{QM}},
\end{align}
where $\Theta $, $\eta $, and $G$ are dimensionless quantities:
\begin{align}
\eta  
&\equiv 
m_{\mathrm{YM}} \ \tau \\
\Theta  
&\equiv 
\sqrt{\frac{2m_{\mathrm{YM}} C V}{a}} \theta \\
G 
&\equiv  
\frac{N}{l_0} \sqrt{\frac{a}{2m_{\mathrm{YM}} C V}}.
\end{align}

Our current goal is to estimate the quantum tunneling amplitude between the two adjacent wells, or two adjacent vacua.
Supposing that $\ket{\mathcal{E}}_+$ and $\ket{\mathcal{E}}_-$ represent states at energy $\mathcal{E}$ localized at $\Theta = +\pi/G$ and $\Theta = - \pi/G$ respectively, the amplitude of the transition in imaginary time $T$ is written as
\begin{align}
A_{+, -} = _+\braket{\mathcal{E}|e^{-2T \hat{H}}|\mathcal{E}}_- = e^{-2\mathcal{E} T} \sinh(2 \Delta \mathcal{E} T).
\label{A1}
\end{align}
 $\Delta \mathcal{E}$ is the off-diagonal component of Hamiltonian employing the states $\{ \ket{\mathcal{E}}_+ \ \ket{\mathcal{E}}_-\}$ as its basis, i.e.
\begin{align}
\begin{bmatrix}
_+\braket{\mathcal{E}|H|\mathcal{E}}_+ 
&_+\braket{\mathcal{E}|H|\mathcal{E}}_- \\
_-\braket{\mathcal{E}|H|\mathcal{E}}_+  
& _-\braket{\mathcal{E}|H|\mathcal{E}}_-
\end{bmatrix}
\equiv
\begin{bmatrix}
\mathcal{E} & - \Delta \mathcal{E} \\
- \Delta \mathcal{E}  & \mathcal{E} 
\end{bmatrix} .
\end{align}

On the other hand, the transition amplitude can be also expressed by path integral:
\begin{align}
A_{+, -}^{(1)} = \int_{\Theta (\eta_i)}^{\Theta (\eta_f)} \mathcal{D} \Theta  \exp \left[- \int_{\eta _i}^{\eta _f} d \tau \ \mathcal{L}_{\Theta }\right],
\end{align}
where $\eta_f - \eta_i \equiv 2T$.
The path integral method is a powerful tool when evaluating the non-perturbative effects, where we can treat the tunneling as quantum fluctuations around the instanton between two wells.

Instanton solution $\Theta _c (\eta )$ have already been exactly calculated in~\cite{1995Liang}:
\begin{align}
\Theta _c (\eta ) = \frac{2}{G} \arcsin (k \ \mathrm{sn}(\eta  - \eta _0, k^2)),
\end{align}
where $\mathrm{sn}(x, k^2)$ is the\textit{ Jacobian elliptic function} with modulus $k \equiv \sqrt{1 - G^2 \mathcal{E}/2}$ and $\eta _0$ denotes the position of the instanton. For zero energy ($\mathcal{E} = 0$), this turns to be 
\begin{align}
\Theta _c (\eta ) = \frac{2}{G} \arcsin (\tanh (\eta  - \eta _0)).
\end{align}

We now set $\Theta (\eta ) = \Theta c (\eta ) + \delta \Theta (\eta )$ and perform the path integral for an exponentiated quadratic form of $\delta \Theta  (\eta )$. 
As shown in~\cite{1995Liang}, performing the path integral over the fluctuation around the classical solution $\bar{\theta_c}$ yields the transition amplitude
\begin{align}
A_{+,-}^{(1)} = \frac{2T}{4 \mathcal{K}(k')} e^{-W} e^{-2 \mathcal{E} T},
\end{align}
where
\begin{align}
W = \frac{8}{G^2} [E(k) - k'^2 \mathcal{K}(k)].
\end{align}
Here, $\mathcal{K}(x)$ and $E(x)$ are the \textit{complete elliptic integrals} of the first and second kinds, respectively, and $k'$ is defined as $\sqrt{1-k^2}$.

Note that the corresponding classical trajectories include not only the single instanton configuration but also multi-instanton configurations such as instanton--anti-instanton--instanton. 
Therefore, the total amplitude is the sum of all the possible configurations:
\begin{align}
A_{+, -} = \sum^{\infty}_{m=0} A_{+, -}^{(2m+1)} 
= e^{-2\mathcal{E} T} \sinh \left( \frac{T}{2 \mathcal{K}(k')} e^{-W}\right).
\label{A2}
\end{align}
Combining (\ref{A1}) and (\ref{A2}) and assuming $\mathcal{E} \simeq 0$, we obtain
\begin{align}
\Delta \mathcal{E} &= \frac{1}{4 \mathcal{K}(k')} e^{-W}
\nonumber \\
&\propto\exp \left[ - \frac{8\sqrt{2}V}{a} \left( l_0^{N+2} e^{-\beta \sigma a} \frac{b_N}{b}\right)^{1/2}\right].
\end{align}
Therefore, as $V \rightarrow \infty$, we obtain
\begin{align}
\begin{bmatrix}
\mathcal{E} & -\Delta \mathcal{E} \\
-\Delta \mathcal{E}  & \mathcal{E} 
\end{bmatrix} 
\longrightarrow 
\begin{bmatrix}
\mathcal{E} & 0\\
0 & \mathcal{E} 
\end{bmatrix} ,
\end{align}
which means that the Hamiltonian becomes diagonal and the different vacua become irreducible.

\bibliographystyle{spphys}
\nocite{*}
\bibliography{main_0616.bib}

\end{document}